\documentclass[journal=nalefd,manuscript=letter]{achemso}

\usepackage{xcolor}
\usepackage{hyperref}
\usepackage[version=3]{mhchem}
\usepackage[T1]{fontenc}
\usepackage{color}

\author{Zhibin Gao}
\affiliation{Center for Phononics and Thermal Energy Science and
	School of Physics Science and Engineering, Tongji University,
	200092 Shanghai, People's Republic of China}

\author{Xiao Dong}
\affiliation{Center for High Pressure Science and Technology
	Advanced Research, Beijing 100193, China}
\email{xiao.dong@hpstar.ac.cn}

\author{Nianbei Li}
\affiliation{Center for Phononics and Thermal Energy Science and
	School of Physics Science and Engineering, Tongji University,
	200092 Shanghai, People's Republic of China}

\author{Jie Ren}
\email{Xonics@tongji.edu.cn}
\affiliation{Center for Phononics and Thermal Energy Science and
	School of Physics Science and Engineering, Tongji University,
	200092 Shanghai, People's Republic of China}

\title[Novel 2D silica with NPR]
{Novel Two-Dimensional Silicon Dioxide with  {in-plane} Negative Poisson's Ratio}


\begin{document}

\begin{abstract}
	
Silicon dioxide or silica, normally existing in various bulk crystalline and amorphous forms, is
recently found to possess a two-dimensional structure. In this work, we use \emph{ab initio} calculation and evolutionary algorithm to unveil three 
new 2D silica structures whose themal, dynamical and mechanical stabilities are compared with many typical 
bulk silica. In particular, we find that all these three 2D silica have large  {in-plane} negative
Poisson's ratios with the largest one being double of penta-graphene and three times of borophenes. The negative Poisson's
ratio originates from the interplay of lattice symmetry and Si-O tetrahedron symmetry. Slab silica
is also an insulating 2D material, with the highest electronic band gap ($>$ 7 eV) among reported 2D structures.
These exotic 2D silica with  {in-plane} negative Poisson's ratios and widest band gaps are expected to have great potential
applications in nanomechanics and nanoelectronics.

\end{abstract}

\section{Introduction}

Silicon dioxide with the chemical formula \ce{SiO2} is the fundamental component of glass, sand and
most minerals, which is also known as one of the building units of earth crust and mantle.
There are a large number of isomers with the formula of \ce{SiO2} in crystalline and amorphous forms,
such as quartz, cristobalite and glassy \ce{SiO2}. In all of the known \ce{SiO2} compounds, Si has
\emph{sp$^3$} hybridization with 4-fold tetrahedron configuration. O is coordinated with 2 Si atoms,
but sometimes with angle type like O in \ce{H2O}, sometimes with linear type like C in \ce{CO2}. The
variety of \ce{SiO2} phases mainly originates from the space stacking form of Si-O tetrahedron. Normally,
bulk silicon dioxides are hard ($\alpha$-quartz with theoretical Vickers hardness 30.2 GPa\cite{hardness}),
high-quality electrical insulators and favorable dielectric materials. Due to the outstanding mechanic
and electronic performances, \ce{SiO2} films and slabs have significant applications in mechanics, optics
and electronics.

Two-dimensional (2D) materials are substances with a monolayer or few atomic layers thickness \cite{Geim,Tomas}.
Because of quantum confinement effect, electrons in these materials only have freedom in 2D plane, which
could give rise to new physics. But so far, the 2D materials are quite rare and most of them are semimetals
(e.g., graphene) or semiconductors (e.g., black phosphorus, sulfide and selenide), with few being insulators,
e.g., h-BN, the reported insulating 2D structure with the widest band gap ( 4.7 eV theoretically \cite{BNgap11,nnano,BNgap1} and 6.4 eV experimentally \cite{BNgap2}). It is well known that 3D silicon dioxide is a good insulator with
wide band gap. Therefore, it is interesting to explore whether the corresponding 2D counterpart may also
be stable as a monolayer and exhibit novel properties. If confirmative, it is likely to make contributions to
transistors of nanomaterials and multifunctional van der Waals (vdW) heterostructures \cite{Yuan,Novoselov}.

Thin silica film has firstly been grown on metal Mo(112) substrate \cite{WeissPRL,Todorova,Schroeder,SeifertPRL}.
Then slab crystalline silica sheet is grown on another metal Ru(0001) \cite{LoffPRL,HeydeCPL}.
Furthermore, this hexagonal quasi-2D silica can even be supported by graphene \cite{HuangNanoLett}.
 {In this letter, we focus on the intrinsic mechanical porperties and use theoretical
structure prediction to explore 2D crystalline freestanding silicon dioxide. We identify three novel stable 2D-silica
phases, besides the reported 2D-silica\cite{LoffPRL}. They all display in-plane negative Poisson's ratio (NPR), large band gaps,
superhard mechanical properties compared to 3D $\alpha$-quartz.}

\section{Computational details}

Crystal structure searching were performed via the $\emph{ab initio}$ evolutionary algorithm
in USPEX \cite{USPEX1,USPEX2,USPEX3} which has been successfully applied to a wide range of
searching problems \cite{Mannix1513,Niu,Dongxiao}.  {Furthermore, we have reconfirmed the global four 
minimum free energy of our structures based particle-swarm optimization as implemented in CALYPSO code \cite{calypso}.} 
We sequentially used 1-6 times of chemical formula \ce{SiO2} (up to 18 atoms) per unit cell for searches. According to the experimental results \cite{WeissPRL,LoffPRL}, the thickness of
hexagonal 2D-silica is 4.34 {\AA}. Therefore, in order to cover some extraordinary 2D-silica,
the original thickness was set up to 5 {\AA} and permitted to relax during the evolution.
The self-consistent energy calculations and structure optimization were employed using the
Perdew-Burke-Ernzerhof (PBE) exchange-correlation functional \cite{PBE} along with the
projector-augmented wave (PAW) potentials \cite{PAW1,PAW2} as implemented in the Vienna
\emph{Ab-initio} Simulation Package (VASP) \cite{VASP1,VASP2}. The kinetic energy cut-off
was 800 eV and tetrahedron method with Bl\"{o}chl correction \cite{Blochl} was used to integrate
the Brillouin-zone. Energy convergence value in self-consistent field (scf) loop was selected
as 10$^{-8}$ eV and a maximum Hellmann-Feynman forces is less than 0.1 meV/{\AA}. Such a high
criteria is found to be required to achieve convergence for phonon calculations and all the
elastic-constant components. All the results in this paper is also checked with the local density
approximation (LDA) functional \cite{LDA}. Additionally, since both LDA and PBE approaches usually
underestimate the band gap of semiconductor and insulator, we adopt the screened hybrid
functional of Heyd, Scuseria, and Ernzerhof (HSE06) \cite{HSE06} for a more accurate calculation.
Phonon dispersion curves were obtained using the Phonopy package \cite{phonopy}.

\section{Results and discussion}

\subsection{Structures}

A number of low-energy structures had been searched and only the most stable four 2D structures 
obtained in our computations are discussed here,  with space groups of \emph{P}6/\emph{mmm} \ce{Si4O8},
\emph{Pbcm} \ce{Si4O8}, \emph{P}-4\emph{m}2 \ce{Si4O8} and \emph{P}-4\emph{m}2 \ce{SiO2}. For simplicity, 
we define them as $\alpha$-, $\beta$-, $\gamma$- and $\delta$-2D-silica, respectively. The optimized four
silica structures were shown in Figure {\ref{fig:Structures}. $\alpha$-2D-silica is hexagonal and has been
experimentally grown on Mo(112) and Ru(0001) surfaces\cite{WeissPRL,LoffPRL}, while we obtain it intrinsically
in a freestanding form.  {This known structure has been discovered in our crystal structure prediction, which does validate the correctness of our computational method in high-throughout computing.} Other three are our new discovered structures during the explorations. $\alpha$-2D-silica has a perfect \emph{sp}$^{3}$ bonding network which means all O atoms are connected to Si atoms with same
solid angles 109$^{\circ}$28$^{\prime}$, i.e., Si is in a perfect O tetrahedron. The lattice constant in 
our computation of $\alpha$-2D-silica is 5.31 {\AA} which is in good agreement with the experimental 
data 5.2-5.5 {\AA} \cite{WeissPRL,LoffPRL}. This confirms the accuracy of our structure prediction and relaxation.

The optimized structural parameters of four structures are summarized in Table {\ref{tab:structure}.
In all of 2D silica, each silicon is connected with four oxygens, constituting a tetrahedron like bulk SiO$_{2}$.
Therefore, silicon atoms share
fourfold coordinations and oxygen two, which means all atoms have no redundant valence electrons. In the
top view, $\alpha$-2D-silica, like graphene, indeed has a hexagonal crystal structure, while other
three have rectangular crystal configurations. Specifically, $\beta$-2D-silica is an anisotropic
structure in orthorhombic lattice while $\gamma$-2D-silica and $\delta$-2D-silica are isotropic in tetragonal lattice.
In the side view, $\alpha$-2D-silica with mirror symmetry to z=0 plane consists of a 'double' layers
\ce{Si2O3} substructures connected by intermediate Si-O-Si bonds oriented perpendicular to the
surfaces. However, other three structures have large inherent bucklings in their interlaminations
and lost mirror symmetry relative to x-y plane at z=0.  $\alpha$-2D-silica and $\delta$-2D-silica each
have only one uniform length of Si-O bond, while $\beta$-2D-silica and $\gamma$-2D-silica have three
different Si-O bond lengths. In top view, the Si-O bonds are like crossroads in $\beta$-2D-silica,
$\gamma$-2D-silica and $\delta$-2D-silica crystal structures.

\begin{figure*}
	\includegraphics[width=15cm]{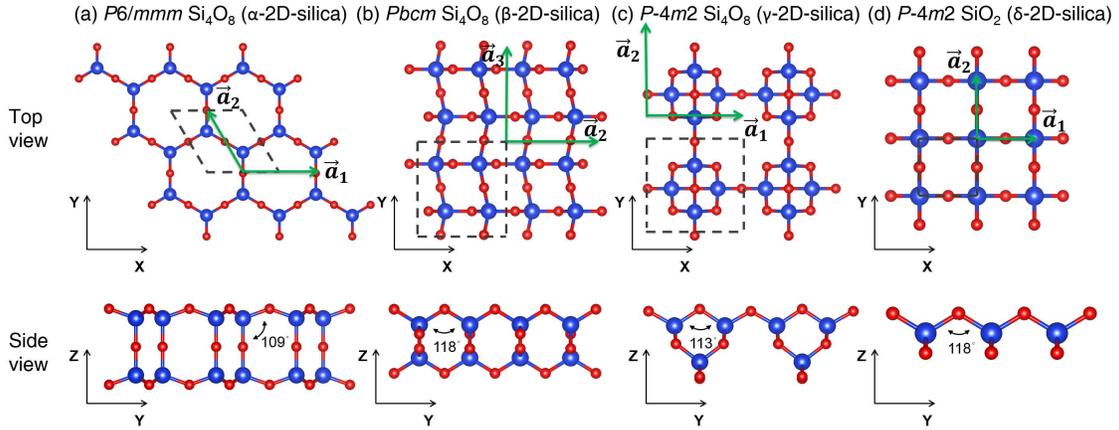}
	\vspace{0.0cm} \caption{\label{fig:Structures}
		(color online). Equilibrium 2D monolayer silicon dioxide of (a) \emph{P}6/\emph{mmm} \ce{Si4O8} ($\alpha$-2D-silica),
		(b) \emph{Pbcm} \ce{Si4O8} ($\beta$-2D-silica), (c) \emph{P}-4\emph{m}2 \ce{Si4O8} ($\gamma$-2D-silica)
		and (d) \emph{P}-4\emph{m}2 \ce{SiO2} ($\delta$-2D-silica) in both top and side views. Large blue and small red
		balls in the (a), (b), (c), (d) represent silicon and oxygen atoms, respectively. The Wigner-Seitz cells are
		shown by the dotted black regions. The green solid arrows show the lattice vectors.
	}
\end{figure*}	
	
The four phases of 2D silica have different arrangement of Si-O tetrahedrons. By
geometry (directions of Si-O bonds consistent with lattice vectors, lattice types and whether have
mirror symmetry relative to z=0 plane), the above four structures can be divided into two classes,
one is $\alpha$-2D-silica, the other are $\beta$-, $\gamma$- and $\delta$-2D-silica. The latter
class has special crossed orientation of chemical bonds which is seldom found in silicon dioxide
crystalline forms.

A deeper insight is that Si-O tetrahedron of symmetry group T$_d$ has two special orthogonal projections:
one goes through a vertex and the center of its opposite face (3-fold axis);
the other goes through on opposite edges (2-fold axis).
For most bulk \ce{SiO2} phases and $\alpha$-2D-silica, the lattice vector is along 3-fold axis, which means
this direction is easy to build a spacial framework by Pauling's third rule \cite{pauling} (rule of sharing of polyhedron corners, edges and faces). The representative local structure is double tetrahedron
structure of [Si$_2$O$_7$]$^{-6}$, which is the fundamental structure of $\alpha$-2D-silica and quartz. Here when the structure
is limited to two dimension, the expansion in three dimensional space become not very important, and the lattice vector
can build the slab along 2-fold axis to get $\beta$-, $\gamma$- and $\delta$-2D-silica. These coupling of
lattice symmetry and Si-O tetrahedron symmetry also is the origin of NPR which will be discussed later.

As quasi-2D structures, the atomic thickness of $\alpha$-, $\beta$-, $\gamma$- and $\delta$-2D-silica
are 4.34, 3.51, 3.84 and 1.71 in {\AA}, respectively. If we add two van der Waals (vdW) radii of the outmost surface
atoms \cite{Baowen}, we can get their effective thicknesses are 7.38, 6.55, 6.88 and 4.75 in {\AA}, respectively.
It is observed that $\delta$-2D-silica is the thinnest structure, while the thickness increases by the sequence of $\delta$,
$\beta$, $\gamma$ and $\alpha$-2D-silica. $\beta$-, $\gamma$- and $\delta$-2D-silica are thinner than $\alpha$-2D-silica,
which implies they are more like pure two-dimensional structures and have more low dimensional effects. Lower thickness
makes these three phases to lose mirror symmetry and they have not enough spaces for double tetrahedron structure of [Si$_2$O$_7$]$^{-6}$,
that is why they have different Si-O bond orientations. In this way, $\beta$-, $\gamma$- and $\delta$-2D-silica can be
seen as a metastable intermediate products when the atoms build silica from zero thickness to $\alpha$-2D-silica. \vspace{0.3cm}	
		
\begin{table*}\scriptsize
	\parbox{1.0\textwidth}{
		\caption
		{Optimized structural properties of the four different 2D silicon dioxide crystal structures.
			$\left|\vec{a}_{1}\right|$, $\left|\vec{a}_{2}\right|$ and $\left|\vec{a}_{3}\right|$
			are the lattice constants defined in Figure 1. Atomic positions are	the group Wyckoff
			positions for each independent atoms in fractional coordinates. \textit{E$_{c}$} is the
			cohesive energy per atom in eV. \textit{E}$_{G-HSE}$ is the band gap calculated by HSE06.
		}
		 \label{tab:structure}}	
		\renewcommand\arraystretch{1.5}
	\begin{tabular*}{1.0\textwidth}{p{2.5cm}*{4}{p{0.192\textwidth}}}
			\hline \hline
			Phase & $\alpha$-2D-silica(\ce{Si4O8}) & $\beta$-2D-silica(\ce{Si4O8}) & $\gamma$-2D-silica(\ce{Si4O8}) & $\delta$-2D-silica(\ce{SiO2})\\
			space group & \emph{P}6/\emph{mmm} & \emph{Pbcm} & \emph{P}-4\emph{m}2 & \emph{P}-4\emph{m}2 \\
			\hline
			$\left|\vec{a}_{1}\right|$({\AA}) & 5.31 & 20.00  & 5.64  & 2.84 \\
			$\left|\vec{a}_{2}\right|$({\AA}) & 5.31 & 5.07 & 5.64  & 2.84 \\
			$\left|\vec{a}_{3}\right|$({\AA}) & 20.00 & 5.53 & 20.00  & 20.00 \\
			\hline
			Atomic Positions & Si 4h (0.67,0.33,0.42) & Si 4d (0.55,0.70,0.25)& Si 4k (0.50,0.75,0.55)& Si 1d (0.00,0.00,0.50) \\

			& O  6i (0.50,0.50,0.39) & O  4d (0.52,0.39,0.25)  & O  2f (0.50,0.50,0.40) & O  2g (0.50,0.00,0.46) \\
			& O  2d (0.33,0.67,0.50) & O  4c (0.59,0.75,0.50)  & O  2g (0.50,0.00,0.59) &    \\
			&                        &                         & O  4i (0.73,0.27,0.50) &    \\
   \textit{E$_{c}$}(eV/atom)         & 6.61                    & 6.54                   & 6.44   & 6.37   \\
		  \textit{E}$_{G-HSE}$(eV)   &  7.31  &  7.69  &  7.40  &  7.34  \\
thickness(\AA) & 4.34 & 3.51 & 3.84 & 1.71\\
			\hline \hline
	\end{tabular*}
\end{table*}

\subsection{Stability and wide bandgap}

\begin{figure}
	\includegraphics[width=0.725\columnwidth]{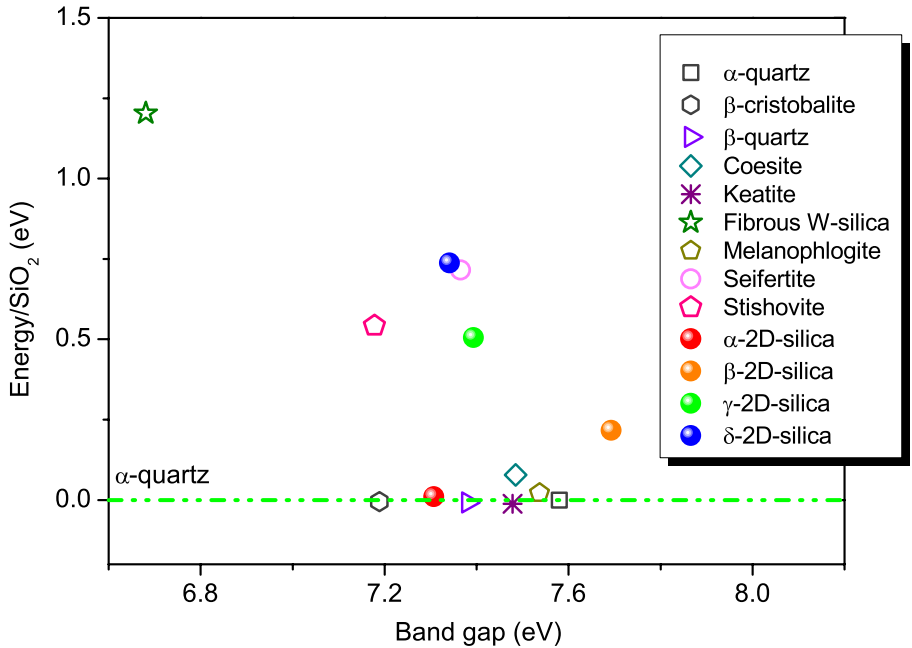}
	\vspace{0.125cm} \caption{\label{fig:energy_landscape}		
		(color online). Energy landscape versus band gap of 3D and 2D silica
		materials as performed in the HSE06 method. As $\alpha$-quartz is the
		most stable silica at low temperatures and is the most common form of
		crystalline silica, we set it as the benchmark reference in green dashed line
		to compare all silica structures.
	}
\end{figure}

Now our first priority is to answer the question whether these new freestanding 2D silicon dioxide
materials are stable. In normal condition, silica exists in many crystalline forms. 
Here we compare the free energies of our 2D silica structures and 3D counterparts in thermodynamics and the calculation are implemented in HSE06 method (Figure {\ref{fig:energy_landscape}}). Compared with $\alpha$-quartz,
who has the lowest energy to be the thermal-dynamically stable phase of \ce{SiO2}, all 2D silica
are meta-stable phases. The structure of $\alpha$-2D-silica with highest symmetry has the lowest
energy (0.047 eV/f. u.) in 2D silica whose energy is comparable to $\alpha$-quartz which is chosen as the reference point of zero energy.
Compared with other 2D-silica, $\alpha$-2D-silica is thicker and
more like bulk silica. That is why it has a similar energy to bulk quartz. Si-O bond length in $\alpha$-2D-silica is 1.63 {\AA} with a quite similar to 1.61 {\AA} of $\alpha$-quartz,
while the Si-O bonds are 1.64 {\AA}, 1.65 {\AA}, 1.66 {\AA} for $\beta$-, $\gamma$- and $\delta$-2D-silica, respectively,
which means there is a positive correlation between energy of structure and Si-O bond length since shorter
distance intensifies the strength of Si-O bonds. In a word, [\ce{SiO4}] tetrahedral unit in $\alpha$-2D-silica
is more similar to $\alpha$-quartz than $\beta$-, $\gamma$- and $\delta$-2D-silica. 

Besides, according to
Pauling's fifth rule (parsimony rule \cite{pauling}), the number of essentially different kinds of constituents in a crystal tends to be
small. The repeating units will tend to identical because each atom in the structure is most stable in
a specific chemical environment. $\alpha$-2D-silica and $\alpha$-quartz have only one type of tetrahedral, while
$\beta$-, $\gamma$- and $\delta$-2D-silica have two or three distorted tetrahedral because of symmetry breaking.
Therefore, the energy of $\alpha$-2D-silica is close to bulk material with lower energy.

 {Though $\alpha$-2D-silica has been grown exprimentally \cite{WeissPRL, SeifertPRL, LoffPRL}, it is not useless to explore other interesting phases with a little higher energy in complicated but brand new 2D silica energy surface. As a matter of fact, metastable phases are common in condensed matter and the ubiquitous metastable phases do not produce a bad effect on their vast and tremendous applications in modern industrial societies and our daily life.} The energies of $\beta$-2D-silica (0.253 eV/f. u.), $\gamma$-2D-silica (0.542 eV/f. u.) and $\delta$-2D-silica (0.758 eV/f. u.)
are comparable to stishovite \cite{stishovite1,stishovite2} and seifertite \cite{Seifertite1}, much
lower than energy of fibrous W-silica \cite{fibrous}. Since stishovite and seifertite are observed by
high pressure experiments and quenchable to ambient condition, and W-silica can be synthesized by chemical
methods, it is reasonable to believe that our three novel 2D phases can exist as a metastable phases at normal condition.
It is also worth highlighting that $\beta$-2D-silica has both the largest band gap (up to 7.69 eV) in
our HSE06 calculations and a moderate low free energy.

The dynamical stability has been confirmed by calculating phonon dispersion relations shown in the
Figure S1. All the structures of four 2D silica are free from imaginary frequencies, which means
they are dynamically stable. Furthermore, the dynamical stability of these freestanding
2D silica is also validated by performing \emph{ab initio} molecular dynamics simulations using
canonical ensemble at a series of elevated temperatures with lifetime longer than 10 ps (Figure S2). The result shows that all 2D-silica are still robust even at high 
temperature (1000 K). Particularly, $\beta$-2D-silica can live in the high temperature of 2500 K which is much higher than the melting point of quartz (1986 K) \cite{deer1992}.

For 2D elastic solid materials, mechanical stability is indispensable to the existence of materials.
If a 2D material is mechanically stable, the corresponding elastic constants have
to satisfy C$_{11}$C$_{22}$-C$_{12}^{2}$$>$0 and C$_{66}$$>$0 \cite{ZhangPNAS}. We have guaranteed
that all the elastic constants (shown in the Table S1) of four 2D silica always
meet this compulsory requirement. Therefore, we have verified that $\alpha$-, $\beta$-, $\gamma$- and $\delta$-2D-silica
are metastable phases.

Band gap is the fundamental electronic property for semiconductor and insulator. Theoretically,
we find that 2D-silica is of the largest band gap in all reported 2D crystal materials.
As the electronic band structures are shown in the Figure S3, $\alpha$-, $\beta$-, $\gamma$-2D-silica
are direct insulators while $\delta$-2D-silica is an indirect insulator. Compared with h-BN which
is reported to have a large band gap of 4.7 eV theoretically \cite{BNgap11,BNgap1}
(6.4 eV experimentally \cite{BNgap2}), silica put the upper limit of 2D band gap to 7.69 eV. This implies it is hard to change its electronic states in 2D-silica. Since the absorption of light is related to the band gap and the thickness, 2D-silica can be the most transparent materials in the world.

As good insulators, 2D-silica can be an outstanding candidate for dielectric materials. Dielectric constants of $\alpha$-, $\beta$-, $\gamma$-2D-silica and $\delta$-2D-silica are 3.18, 2.42, 2.24 and 2.49, respectively. Notably, based on the formula of in-plane parallel capacitor $ c= \varepsilon S/4 \pi kd $, capacitance of $\delta$-2D-silica is around 1.76 times larger than $\alpha$-quartz because it is 0.31 times thinner than $\alpha$-quartz. This means it is a favorable dielectric material for next nanomaterials transistors such as 2D transistors and 2D electronic devices \cite{Franklin} and a supercapacitor candidate \cite{bonaccorso}.

\subsection{Superhard mechanical properties with NPR}
	
When a material is stretched in one direction, it usually contracts in the other two directions
perpendicular to the applied stretching direction, so-called positive Poisson's ratio (PPR). Most materials have a PPRs 
ranging from 0 to 0.5. From the classical theory of elasticity, for 3D isotropic materials,
the Poisson's ratio cannot be less than -1.0 or greater than 0.5 \cite{Landau}.
Materials with NPR have the property that when stretched, they become auxetic perpendicular
to the applied force. These materials are also called auxetics by its auxetic property with
external strain. This interesting phenomenon was firstly presented in foam structure in 1987 \cite{Lakes}.
 {And then materials with NPR attract a lot of scientific interest by its wide applications, such as single-layer graphene and graphene ribbons \cite{JiangG1,JiangG2}, single-layer black phosphorus \cite{Jiang,DuBP,jiang2016auxetic}, h$\alpha$-silica \cite{ozccelik2014stable}, penta-graphene \cite{ZhangPNAS}, borophenes \cite{Mannix1513}, and semi-metallic \ce{Be5C2} \cite{wang2016semi}, and can be used in vanes for aircraft \cite{baughman1998}, packing materials \cite{grima2006}, body armor and national defense \cite{liu2006auxetic}. Furthermore, we find that in-plane NPR is rare and novel in 2D materials based on the collected data from the available published papers with our utmost endeavor shown in the Figure S4.} \vspace{0.3cm} 												
\begin{figure}
	\includegraphics[width=0.735\columnwidth]{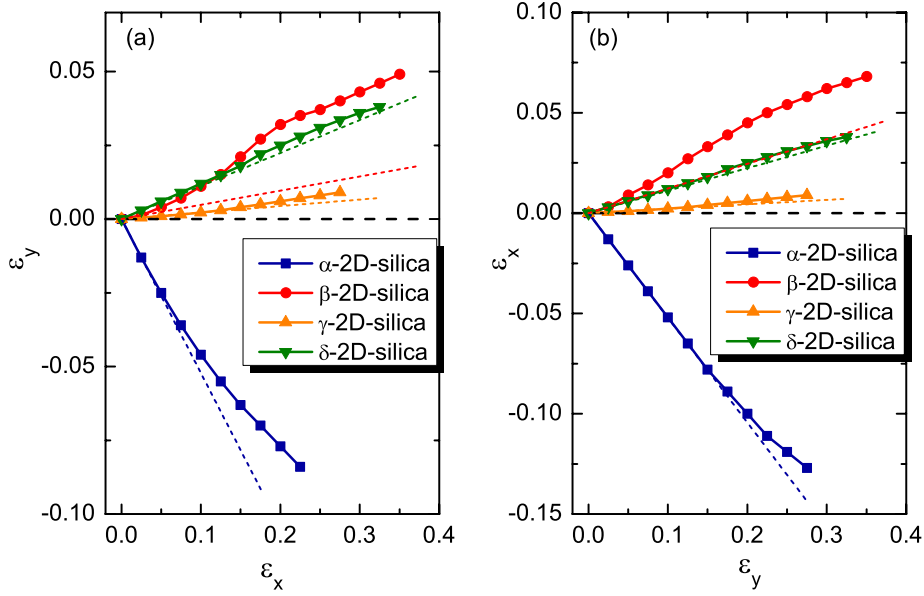}
	\vspace{0.175cm} \caption{\label{fig:poission_ratio}	
	(color online). The Poisson's ratio (short dashed lines denote the linear response counterpart) 
	as a function of uniaxial tensile strain of $\alpha$-2D-silica, $\beta$-2D-silica, $\gamma$-2D-silica and $\delta$-2D-silica
	in the (a) x-direction ($\nu_{xy}$= 0.522, -0.048, -0.022 and -0.112) and
	(b) y-direction ($\nu_{yx}$= 0.522, -0.123, -0.022, -0.112). The horizontal
	black dashed lines are set to zero for comparison.
	}
\end{figure}

Poission's ratio $\nu$ and Young's modulus $\textit{E}$ in the xy plane can be expressed as following equations:\vspace{-0.5cm}
\begin{equation}\label{mechanicals1}
	\begin{split}
	& \nu_{xy}=-\frac{d\varepsilon_{y}}{d\varepsilon_{x}}, {~~~~~~} \nu_{yx}=-\frac{d\varepsilon_{x}}{d\varepsilon_{y}}, \\
	& E_{x}=\frac{\sigma(\varepsilon_{x})}{\varepsilon_{x}}, {~~~~~~} E_{y}=\frac{\sigma(\varepsilon_{y})}{\varepsilon_{y}}, \\
	\end{split}
\end{equation}			
Where $\nu_{xy}$ is the Poisson's ratio of the x direction which is induced by a strain
in the x axis. $\textit{E$_{x}$}$ is the Young's modulus of the x direction which is the slope in the
stress-strain curve. Similarly, $\nu_{yx}$ and $\textit{E$_{y}$}$ are the Poisson's ratio and Young's
modulus in the y direction. $\sigma$ is the stress as a function of strain $\varepsilon$. We quantitatively
calculated the Poisson's ratio $\nu$ and Young's modulus \textit{E} for our four 2D silicon dioxides according to
the Eq. \eqref{mechanicals1} for uniaxial strain. The results of Poisson's ratio are shown in Figure {\ref{fig:poission_ratio}}.
We also calculated the mechanical property by the elastic solid theory (the detailed description of this
method is discussed in the Supporting Information). The results of elastic solid theory and strain method are
quite similar(Table S2). Therefore, both methods confirm that our three 2D silica phases have NPR mechanical property.
In order to check the correctness of our computational method and results, we select graphene and black phosphorus
as references. Our results shown in the Table S2 are in good agreement with the published results \cite{WeiPRB,DanielPRB,Qunwei,JiweiNatcom}.
As we have mentioned above, $\alpha$-2D-silica, $\gamma$-2D-silica and $\delta$-2D-silica
are isotropic materials, while $\beta$-2D-silica is an anisotropic material. Therefore, except $\beta$-2D-silica, all other
three materials have the mechanical properties \textit{E}$_{x}$=\textit{E}$_{y}$ and $\nu_{x}$=$\nu_{y}$.

Figure {\ref{fig:poission_ratio}}(a) displays the strain response in the y direction when applying stretch
in the x direction. Similarly, the Poisson's ratio of the y direction is shown in Figure {\ref{fig:poission_ratio}}(b).
Obviously, there are two kinds of mechanical response when four 2D silica are stretched. $\alpha$-2D-silica is
contracted perpendicular to the applied force, while $\beta$-2D-silica, $\gamma$-2D-silica and $\delta$-2D-silica
are auxetic during the uniaxial strain in 2D plane. Remarkably, $\beta$-2D-silica in the y-direction and $\delta$-2D-silica
in both x and y directions have large NPR values -0.123 and -0.112. These absolute values of NPR are
quite high, which equal to double of penta-graphene ($\nu$=-0.068) \cite{ZhangPNAS}, three times of
borophenes ($\nu_{x}$=-0.04, $\nu_{y}$=-0.02) \cite{Mannix1513}, and are different from single-layer
black phosphorus whose NPR occurs in the yz plane \cite{Jiang}.

\begin{figure}
	\includegraphics[width=0.55\columnwidth]{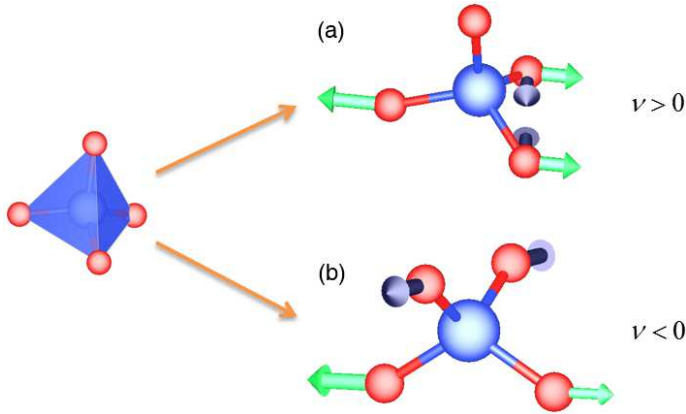}
	\vspace{0.2cm}\caption{\label{fig:explainNPR}
	(color online). The explanation of Poisson's ratio for our new 2D silicon
	dioxide structures. Each silicon atom is surrounded by four oxygen atoms,
	due to its shape, the \ce{SiO4} groups is called \ce{SiO4} tetrahedron.
	(a) when the tensile strain (in green arrows) is applied to the direction of
	the median line of bottom of the tetrahedron (vertical to 3-fold axis), because the angular variation,
	the distance of two oxygen atoms decreases (in black arrows), leading a positive
	Poisson's ratio ($\nu$$>$0). On the contrary, (b) when the tensile strain is
	applied to the edge to the tetrahedron (vertical to 2-fold axis), similarly, because the angular variation,
	the distance of two oxygen atoms increases, leading a NPR phenomenon ($\nu$$<$0).
	}
\end{figure}

NPR of our novel three 2D silica structures originates from the low dimensional effect. 
As discussed in structure section, [\ce{SiO4}] tetrahedron is the basic units in both slab and
bulk \ce{SiO2}. Normally in 3D system, by the request of sharing of polyhedron corners (Pauling's third rule \cite{pauling}), the lattice is along 3-fold axis of Si-O tetrahedron. The tensile
strain (in green arrows) on the direction of the median line of bottom of the tetrahedron makes
the distance of two oxygen atoms decrease (in black arrows), which implies a positive Poisson's
ratio ($\nu$$>$0) shown in Figure {\ref{fig:explainNPR}}(a). However, for 2D materials, there is spatial constrains and the lattice can be along in 2-fold axis. When the tensile strain
is applied to the edge to the tetrahedron, the distance of two oxygen atoms increases, leading to a
NPR phenomenon ($\nu$$<$0) shown in Fig. {\ref{fig:explainNPR}}(b). Therefore, we believe the
distinction between these two situations stems from the coupling of lattice symmetry and Si-O
tetrahedron symmetry, which is a low dimensional effect.

In a word, our three new structures are thinner than $\alpha$-2D-silica with stronger low dimensional effect in geometry. This low dimensional effect further leads to a different type of matching between lattice symmetry and Si-O tetrahedron symmetry, which finally leads to this interesting NPR phenomenon \cite{Keskar}.

We have also revealed the Young's modulus of these 2D silica materials and prove they are superhard 2D materials. As we have
explained above, we use the effective thickness of 2D materials, including two vdW radii \cite{XufeiNL,XufeiarXiv,Baowen}.
The calculated Young's modulus are exhibited in the Table S2. Generally, Young's modulus
can be obtained by dividing the tensile stress by the extensional strain in the elastic region of the
stress-strain graph. A material whose Young's modulus is high can be regarded as rigid. According to
the experimental \cite{Zubov,Ohno} and theoretical \cite{Kim} values, the Young's modulus of 3D
$\alpha$-quartz is around 100 GPa without extra external pressure. It is to be disclosed that all
four 2D silica are much harder than $\alpha$-quartz. Young's modulus of $\delta$-2D-silica (346 GPa)
is more than triple of that of $\alpha$-quartz (100 GPa) \cite{Zubov,Ohno,Kim} and is comparable to that of single-layer Boron
Nitride (BN) (366 GPa) \cite{Andrew}. Interestingly, the Young's modulus of $\beta$-2D-silica in the y
direction (247 GPa) is more than double of that in the x direction (97 GPa) which due to the strong anisotropy
of its special crystal structures. The hardness of these 2D silica can be explained by
tetrahedral silica structure in Figure {\ref{fig:explainNPR}} and phonon dispersion realations in the
Figure S1. The slope of phonon dispersions denotes the group velocity of each phonon. The
bigger group velocity, the more rigid of the material. Therefore, $\delta$-2D-silica and $\beta$-2D-silica have a
superhard and strong anisotropic mechanical properties.

\section{Couclusion}

In conclusion, we have performed a systematic structure searching and computationally-guided material discovery for 2D silicon dioxide and found
$\beta$-, $\gamma$- and $\delta$-2D-silica. The thermal, dynamical and mechanical stability checking
guarantees that these 2D-silica can exist as metastable phases at atmosphere condition. In particular,
 {in-plane NPR} has been identified in these three structures, whose values are so large that are double that of penta-graphene \cite{ZhangPNAS}, three
times of borophenes ($\nu_{x}$=-0.04, $\nu_{y}$=-0.02) \cite{Mannix1513}, and are different from
single-layer black phosphorus's NPR in the yz plane \cite{Jiang}. The NPR originates from the coupling of lattice symmetry and Si-O tetrahedron symmetry which is a low dimensional effect. We have also proved these three 2D silica to be superhard materials, which indicate superior thermal conductivities compared to bulk silica materials. Furthermore, with its largest electronic band gaps in reported 2D crystal materials, we believe these novel 2D-silica can be used as the most transparent insulators and good dielectric materials, which is expected to have great applications in many fields.

The anthors declare no competing financial interest.

Supporting Information Available: Novel Two-Dimensional Silicon Dioxide with  {in-plane} Negative Poisson's Ratio.
This material is available free of charge via the Internet at http://pubs.acs.org. 
							
\section{Acknowledgments}

The numerical calculations were carried out at Shanghai Supercomputer Center. This work is supported by the 
National Youth 1000 Talents Program in China, and the 985 startup grant (No. 205020516074) at Tongji University, (Z.G. and J.R.), and 
by the NSF China with grant No. 11334007(N.L.). X.D. also acknowledges the support of NSAF with grant No. U1530402.


\begin{mcitethebibliography}{66}
\providecommand*\natexlab[1]{#1}
\providecommand*\mciteSetBstSublistMode[1]{}
\providecommand*\mciteSetBstMaxWidthForm[2]{}
\providecommand*\mciteBstWouldAddEndPuncttrue
  {\def\EndOfBibitem{\unskip.}}
\providecommand*\mciteBstWouldAddEndPunctfalse
  {\let\EndOfBibitem\relax}
\providecommand*\mciteSetBstMidEndSepPunct[3]{}
\providecommand*\mciteSetBstSublistLabelBeginEnd[3]{}
\providecommand*\EndOfBibitem{}
\mciteSetBstSublistMode{f}
\mciteSetBstMaxWidthForm{subitem}{(\alph{mcitesubitemcount})}
\mciteSetBstSublistLabelBeginEnd
  {\mcitemaxwidthsubitemform\space}
  {\relax}
  {\relax}

\bibitem[Chen \latin{et~al.}(2011)Chen, Niu, Li, and Li]{hardness}
Chen,~X.-Q.; Niu,~H.; Li,~D.; Li,~Y. \emph{Intermetallics} \textbf{2011},
  \emph{19}, 1275 -- 1281\relax
\mciteBstWouldAddEndPuncttrue
\mciteSetBstMidEndSepPunct{\mcitedefaultmidpunct}
{\mcitedefaultendpunct}{\mcitedefaultseppunct}\relax
\EndOfBibitem
\bibitem[Geim and Grigorieva(2013)Geim, and Grigorieva]{Geim}
Geim,~A.~K.; Grigorieva,~I.~V. \emph{Nature} \textbf{2013}, \emph{499},
  419--425\relax
\mciteBstWouldAddEndPuncttrue
\mciteSetBstMidEndSepPunct{\mcitedefaultmidpunct}
{\mcitedefaultendpunct}{\mcitedefaultseppunct}\relax
\EndOfBibitem
\bibitem[Miro \latin{et~al.}(2014)Miro, Audiffred, and Heine]{Tomas}
Miro,~P.; Audiffred,~M.; Heine,~T. \emph{Chem. Soc. Rev.} \textbf{2014},
  \emph{43}, 6537--6554\relax
\mciteBstWouldAddEndPuncttrue
\mciteSetBstMidEndSepPunct{\mcitedefaultmidpunct}
{\mcitedefaultendpunct}{\mcitedefaultseppunct}\relax
\EndOfBibitem
\bibitem[Peng and De(2012)Peng, and De]{BNgap11}
Peng,~Q.; De,~S. \emph{Physica E} \textbf{2012}, \emph{44}, 1662--1666\relax
\mciteBstWouldAddEndPuncttrue
\mciteSetBstMidEndSepPunct{\mcitedefaultmidpunct}
{\mcitedefaultendpunct}{\mcitedefaultseppunct}\relax
\EndOfBibitem
\bibitem[Churchill and Jarillo-Herrero(2014)Churchill, and
  Jarillo-Herrero]{nnano}
Churchill,~H.~O.; Jarillo-Herrero,~P. \emph{Nat Nanotechnol.} \textbf{2014},
  \emph{9 (5)}, 330 -- 331\relax
\mciteBstWouldAddEndPuncttrue
\mciteSetBstMidEndSepPunct{\mcitedefaultmidpunct}
{\mcitedefaultendpunct}{\mcitedefaultseppunct}\relax
\EndOfBibitem
\bibitem[Xu and Ching(1991)Xu, and Ching]{BNgap1}
Xu,~Y.-N.; Ching,~W.~Y. \emph{Phys. Rev. B} \textbf{1991}, \emph{44},
  7787--7798\relax
\mciteBstWouldAddEndPuncttrue
\mciteSetBstMidEndSepPunct{\mcitedefaultmidpunct}
{\mcitedefaultendpunct}{\mcitedefaultseppunct}\relax
\EndOfBibitem
\bibitem[Zunger \latin{et~al.}(1976)Zunger, Katzir, and Halperin]{BNgap2}
Zunger,~A.; Katzir,~A.; Halperin,~A. \emph{Phys. Rev. B} \textbf{1976},
  \emph{13}, 5560--5573\relax
\mciteBstWouldAddEndPuncttrue
\mciteSetBstMidEndSepPunct{\mcitedefaultmidpunct}
{\mcitedefaultendpunct}{\mcitedefaultseppunct}\relax
\EndOfBibitem
\bibitem[Liu \latin{et~al.}(2016)Liu, Weiss, Duan, Cheng, Huang, and
  Duan]{Yuan}
Liu,~Y.; Weiss,~N.~O.; Duan,~X.; Cheng,~H.-C.; Huang,~Y.; Duan,~X. \emph{Nat.
  Rev. Mater.} \textbf{2016}, \emph{1}, 16042\relax
\mciteBstWouldAddEndPuncttrue
\mciteSetBstMidEndSepPunct{\mcitedefaultmidpunct}
{\mcitedefaultendpunct}{\mcitedefaultseppunct}\relax
\EndOfBibitem
\bibitem[Novoselov \latin{et~al.}(2016)Novoselov, Mishchenko, Carvalho, and
  Castro~Neto]{Novoselov}
Novoselov,~K.~S.; Mishchenko,~A.; Carvalho,~A.; Castro~Neto,~A.~H.
  \emph{Science} \textbf{2016}, \emph{353}\relax
\mciteBstWouldAddEndPuncttrue
\mciteSetBstMidEndSepPunct{\mcitedefaultmidpunct}
{\mcitedefaultendpunct}{\mcitedefaultseppunct}\relax
\EndOfBibitem
\bibitem[Weissenrieder \latin{et~al.}(2005)Weissenrieder, Kaya, Lu, Gao,
  Shaikhutdinov, Freund, Sierka, Todorova, and Sauer]{WeissPRL}
Weissenrieder,~J.; Kaya,~S.; Lu,~J.-L.; Gao,~H.-J.; Shaikhutdinov,~S.;
  Freund,~H.-J.; Sierka,~M.; Todorova,~T.~K.; Sauer,~J. \emph{Phys. Rev. Lett.}
  \textbf{2005}, \emph{95}, 076103\relax
\mciteBstWouldAddEndPuncttrue
\mciteSetBstMidEndSepPunct{\mcitedefaultmidpunct}
{\mcitedefaultendpunct}{\mcitedefaultseppunct}\relax
\EndOfBibitem
\bibitem[Todorova \latin{et~al.}(2006)Todorova, Sierka, Sauer, Kaya,
  Weissenrieder, Lu, Gao, Shaikhutdinov, and Freund]{Todorova}
Todorova,~T.~K.; Sierka,~M.; Sauer,~J.; Kaya,~S.; Weissenrieder,~J.; Lu,~J.-L.;
  Gao,~H.-J.; Shaikhutdinov,~S.; Freund,~H.-J. \emph{Phys. Rev. B}
  \textbf{2006}, \emph{73}, 165414\relax
\mciteBstWouldAddEndPuncttrue
\mciteSetBstMidEndSepPunct{\mcitedefaultmidpunct}
{\mcitedefaultendpunct}{\mcitedefaultseppunct}\relax
\EndOfBibitem
\bibitem[Schroeder \latin{et~al.}(2000)Schroeder, Adelt, Richter, Naschitzki,
  B{\"a}umer, and Freund]{Schroeder}
Schroeder,~T.; Adelt,~M.; Richter,~B.; Naschitzki,~M.; B{\"a}umer,~M.;
  Freund,~H.-J. \emph{Surf. Rev. Lett.} \textbf{2000}, \emph{7}, 7--14\relax
\mciteBstWouldAddEndPuncttrue
\mciteSetBstMidEndSepPunct{\mcitedefaultmidpunct}
{\mcitedefaultendpunct}{\mcitedefaultseppunct}\relax
\EndOfBibitem
\bibitem[Seifert \latin{et~al.}(2009)Seifert, Blauth, and Winter]{SeifertPRL}
Seifert,~J.; Blauth,~D.; Winter,~H. \emph{Phys. Rev. Lett.} \textbf{2009},
  \emph{103}, 017601\relax
\mciteBstWouldAddEndPuncttrue
\mciteSetBstMidEndSepPunct{\mcitedefaultmidpunct}
{\mcitedefaultendpunct}{\mcitedefaultseppunct}\relax
\EndOfBibitem
\bibitem[L\"offler \latin{et~al.}(2010)L\"offler, Uhlrich, Baron, Yang, Yu,
  Lichtenstein, Heinke, B\"uchner, Heyde, Shaikhutdinov, Freund, W\l{}odarczyk,
  Sierka, and Sauer]{LoffPRL}
L\"offler,~D.; Uhlrich,~J.~J.; Baron,~M.; Yang,~B.; Yu,~X.; Lichtenstein,~L.;
  Heinke,~L.; B\"uchner,~C.; Heyde,~M.; Shaikhutdinov,~S.; Freund,~H.-J.;
  W\l{}odarczyk,~R.; Sierka,~M.; Sauer,~J. \emph{Phys. Rev. Lett.}
  \textbf{2010}, \emph{105}, 146104\relax
\mciteBstWouldAddEndPuncttrue
\mciteSetBstMidEndSepPunct{\mcitedefaultmidpunct}
{\mcitedefaultendpunct}{\mcitedefaultseppunct}\relax
\EndOfBibitem
\bibitem[Heyde \latin{et~al.}(2012)Heyde, Shaikhutdinov, and Freund]{HeydeCPL}
Heyde,~M.; Shaikhutdinov,~S.; Freund,~H.-J. \emph{Chem. Phys. Lett.}
  \textbf{2012}, \emph{550}, 1--7\relax
\mciteBstWouldAddEndPuncttrue
\mciteSetBstMidEndSepPunct{\mcitedefaultmidpunct}
{\mcitedefaultendpunct}{\mcitedefaultseppunct}\relax
\EndOfBibitem
\bibitem[Huang \latin{et~al.}(2012)Huang, Kurasch, Srivastava, Skakalova,
  Kotakoski, Krasheninnikov, Hovden, Mao, Meyer, Smet, Muller, and
  Kaiser]{HuangNanoLett}
Huang,~P.~Y.; Kurasch,~S.; Srivastava,~A.; Skakalova,~V.; Kotakoski,~J.;
  Krasheninnikov,~A.~V.; Hovden,~R.; Mao,~Q.; Meyer,~J.~C.; Smet,~J.;
  Muller,~D.~A.; Kaiser,~U. \emph{Nano Lett.} \textbf{2012}, \emph{12},
  1081--1086\relax
\mciteBstWouldAddEndPuncttrue
\mciteSetBstMidEndSepPunct{\mcitedefaultmidpunct}
{\mcitedefaultendpunct}{\mcitedefaultseppunct}\relax
\EndOfBibitem
\bibitem[Oganov and Glass(2006)Oganov, and Glass]{USPEX1}
Oganov,~A.~R.; Glass,~C.~W. \emph{J. Chem. Phys.} \textbf{2006}, \emph{124},
  244704\relax
\mciteBstWouldAddEndPuncttrue
\mciteSetBstMidEndSepPunct{\mcitedefaultmidpunct}
{\mcitedefaultendpunct}{\mcitedefaultseppunct}\relax
\EndOfBibitem
\bibitem[Oganov \latin{et~al.}(2011)Oganov, Lyakhov, and Valle]{USPEX2}
Oganov,~A.~R.; Lyakhov,~A.~O.; Valle,~M. \emph{Acc. Chem. Res.} \textbf{2011},
  \emph{44}, 227--237\relax
\mciteBstWouldAddEndPuncttrue
\mciteSetBstMidEndSepPunct{\mcitedefaultmidpunct}
{\mcitedefaultendpunct}{\mcitedefaultseppunct}\relax
\EndOfBibitem
\bibitem[Lyakhov \latin{et~al.}(2013)Lyakhov, Oganov, Stokes, and Zhu]{USPEX3}
Lyakhov,~A.~O.; Oganov,~A.~R.; Stokes,~H.~T.; Zhu,~Q. \emph{Comput. Phys.
  Commun.} \textbf{2013}, \emph{184}, 1172--1182\relax
\mciteBstWouldAddEndPuncttrue
\mciteSetBstMidEndSepPunct{\mcitedefaultmidpunct}
{\mcitedefaultendpunct}{\mcitedefaultseppunct}\relax
\EndOfBibitem
\bibitem[Mannix \latin{et~al.}(2015)Mannix, Zhou, Kiraly, Wood, Alducin, Myers,
  Liu, Fisher, Santiago, Guest, Yacaman, Ponce, Oganov, Hersam, and
  Guisinger]{Mannix1513}
Mannix,~A.~J.; Zhou,~X.-F.; Kiraly,~B.; Wood,~J.~D.; Alducin,~D.; Myers,~B.~D.;
  Liu,~X.; Fisher,~B.~L.; Santiago,~U.; Guest,~J.~R.; Yacaman,~M.~J.;
  Ponce,~A.; Oganov,~A.~R.; Hersam,~M.~C.; Guisinger,~N.~P. \emph{Science}
  \textbf{2015}, \emph{350}, 1513--1516\relax
\mciteBstWouldAddEndPuncttrue
\mciteSetBstMidEndSepPunct{\mcitedefaultmidpunct}
{\mcitedefaultendpunct}{\mcitedefaultseppunct}\relax
\EndOfBibitem
\bibitem[Niu \latin{et~al.}(2015)Niu, Oganov, Chen, and Li]{Niu}
Niu,~H.; Oganov,~A.~R.; Chen,~X.-Q.; Li,~D. \emph{Sci. Rep.} \textbf{2015},
  \emph{5}, 18347\relax
\mciteBstWouldAddEndPuncttrue
\mciteSetBstMidEndSepPunct{\mcitedefaultmidpunct}
{\mcitedefaultendpunct}{\mcitedefaultseppunct}\relax
\EndOfBibitem
\bibitem[Dong \latin{et~al.}(2016)Dong, Li, Oganov, Li, Zheng, and
  Mao]{Dongxiao}
Dong,~X.; Li,~Y.-L.; Oganov,~A.~R.; Li,~K.; Zheng,~H.; Mao,~H.-k. \emph{arXiv
  preprint arXiv:1603.02880} \textbf{2016}, \relax
\mciteBstWouldAddEndPunctfalse
\mciteSetBstMidEndSepPunct{\mcitedefaultmidpunct}
{}{\mcitedefaultseppunct}\relax
\EndOfBibitem
\bibitem[Wang \latin{et~al.}(2010)Wang, Lv, Zhu, and Ma]{calypso}
Wang,~Y.; Lv,~J.; Zhu,~L.; Ma,~Y. \emph{Physical Review B} \textbf{2010},
  \emph{82}, 094116\relax
\mciteBstWouldAddEndPuncttrue
\mciteSetBstMidEndSepPunct{\mcitedefaultmidpunct}
{\mcitedefaultendpunct}{\mcitedefaultseppunct}\relax
\EndOfBibitem
\bibitem[Perdew \latin{et~al.}(1996)Perdew, Burke, and Ernzerhof]{PBE}
Perdew,~J.~P.; Burke,~K.; Ernzerhof,~M. \emph{Phys. Rev. Lett.} \textbf{1996},
  \emph{77}, 3865--3868\relax
\mciteBstWouldAddEndPuncttrue
\mciteSetBstMidEndSepPunct{\mcitedefaultmidpunct}
{\mcitedefaultendpunct}{\mcitedefaultseppunct}\relax
\EndOfBibitem
\bibitem[Bl\"ochl(1994)]{PAW1}
Bl\"ochl,~P.~E. \emph{Phys. Rev. B} \textbf{1994}, \emph{50},
  17953--17979\relax
\mciteBstWouldAddEndPuncttrue
\mciteSetBstMidEndSepPunct{\mcitedefaultmidpunct}
{\mcitedefaultendpunct}{\mcitedefaultseppunct}\relax
\EndOfBibitem
\bibitem[Kresse and Joubert(1999)Kresse, and Joubert]{PAW2}
Kresse,~G.; Joubert,~D. \emph{Phys. Rev. B} \textbf{1999}, \emph{59},
  1758--1775\relax
\mciteBstWouldAddEndPuncttrue
\mciteSetBstMidEndSepPunct{\mcitedefaultmidpunct}
{\mcitedefaultendpunct}{\mcitedefaultseppunct}\relax
\EndOfBibitem
\bibitem[Kresse and Furthm{\"u}ller(1996)Kresse, and Furthm{\"u}ller]{VASP1}
Kresse,~G.; Furthm{\"u}ller,~J. \emph{Phys. Rev. B} \textbf{1996}, \emph{54},
  11169--11186\relax
\mciteBstWouldAddEndPuncttrue
\mciteSetBstMidEndSepPunct{\mcitedefaultmidpunct}
{\mcitedefaultendpunct}{\mcitedefaultseppunct}\relax
\EndOfBibitem
\bibitem[Kresse and Furthm{\"u}ller(1996)Kresse, and Furthm{\"u}ller]{VASP2}
Kresse,~G.; Furthm{\"u}ller,~J. \emph{J. Comput. Mater. Sci.} \textbf{1996},
  \emph{6}, 15--50\relax
\mciteBstWouldAddEndPuncttrue
\mciteSetBstMidEndSepPunct{\mcitedefaultmidpunct}
{\mcitedefaultendpunct}{\mcitedefaultseppunct}\relax
\EndOfBibitem
\bibitem[Bl\"ochl \latin{et~al.}(1994)Bl\"ochl, Jepsen, and Andersen]{Blochl}
Bl\"ochl,~P.~E.; Jepsen,~O.; Andersen,~O.~K. \emph{Phys. Rev. B} \textbf{1994},
  \emph{49}, 16223--16233\relax
\mciteBstWouldAddEndPuncttrue
\mciteSetBstMidEndSepPunct{\mcitedefaultmidpunct}
{\mcitedefaultendpunct}{\mcitedefaultseppunct}\relax
\EndOfBibitem
\bibitem[Perdew and Zunger(1981)Perdew, and Zunger]{LDA}
Perdew,~J.~P.; Zunger,~A. \emph{Phys. Rev. B} \textbf{1981}, \emph{23},
  5048--5079\relax
\mciteBstWouldAddEndPuncttrue
\mciteSetBstMidEndSepPunct{\mcitedefaultmidpunct}
{\mcitedefaultendpunct}{\mcitedefaultseppunct}\relax
\EndOfBibitem
\bibitem[Heyd \latin{et~al.}(2003)Heyd, Scuseria, and Ernzerhof]{HSE06}
Heyd,~J.; Scuseria,~G.~E.; Ernzerhof,~M. \emph{J. Chem. Phys.} \textbf{2003},
  \emph{118}, 8207--8215\relax
\mciteBstWouldAddEndPuncttrue
\mciteSetBstMidEndSepPunct{\mcitedefaultmidpunct}
{\mcitedefaultendpunct}{\mcitedefaultseppunct}\relax
\EndOfBibitem
\bibitem[Togo \latin{et~al.}(2008)Togo, Oba, and Tanaka]{phonopy}
Togo,~A.; Oba,~F.; Tanaka,~I. \emph{Phys. Rev. B} \textbf{2008}, \emph{78},
  134106\relax
\mciteBstWouldAddEndPuncttrue
\mciteSetBstMidEndSepPunct{\mcitedefaultmidpunct}
{\mcitedefaultendpunct}{\mcitedefaultseppunct}\relax
\EndOfBibitem
\bibitem[Pauling(1960)]{pauling}
Pauling,~L. \emph{The nature of the chemical bond, 3rd ed.}; Cornell University
  Press: Ithaca, NY, 1960\relax
\mciteBstWouldAddEndPuncttrue
\mciteSetBstMidEndSepPunct{\mcitedefaultmidpunct}
{\mcitedefaultendpunct}{\mcitedefaultseppunct}\relax
\EndOfBibitem
\bibitem[Xu \latin{et~al.}(2015)Xu, Zhang, and Li]{Baowen}
Xu,~W.; Zhang,~G.; Li,~B. \emph{J. Chem. Phys.} \textbf{2015}, \emph{143},
  154703\relax
\mciteBstWouldAddEndPuncttrue
\mciteSetBstMidEndSepPunct{\mcitedefaultmidpunct}
{\mcitedefaultendpunct}{\mcitedefaultseppunct}\relax
\EndOfBibitem
\bibitem[Smyth \latin{et~al.}(1995)Smyth, Swope, and Pawley]{stishovite1}
Smyth,~J.~R.; Swope,~R.~J.; Pawley,~A.~R. \emph{Am. Mineral.} \textbf{1995},
  \emph{80}, 454--456\relax
\mciteBstWouldAddEndPuncttrue
\mciteSetBstMidEndSepPunct{\mcitedefaultmidpunct}
{\mcitedefaultendpunct}{\mcitedefaultseppunct}\relax
\EndOfBibitem
\bibitem[Glsplnrr(1990)]{stishovite2}
Glsplnrr,~T. \emph{Am. Mineral.} \textbf{1990}, \emph{75}, 739--747\relax
\mciteBstWouldAddEndPuncttrue
\mciteSetBstMidEndSepPunct{\mcitedefaultmidpunct}
{\mcitedefaultendpunct}{\mcitedefaultseppunct}\relax
\EndOfBibitem
\bibitem[Dera \latin{et~al.}(2002)Dera, Prewitt, Boctor, and
  Hemley]{Seifertite1}
Dera,~P.; Prewitt,~C.~T.; Boctor,~N.~Z.; Hemley,~R.~J. \emph{Am. Mineral.}
  \textbf{2002}, \emph{87}, 1018--1023\relax
\mciteBstWouldAddEndPuncttrue
\mciteSetBstMidEndSepPunct{\mcitedefaultmidpunct}
{\mcitedefaultendpunct}{\mcitedefaultseppunct}\relax
\EndOfBibitem
\bibitem[Weiss and Weiss(1954)Weiss, and Weiss]{fibrous}
Weiss,~A.; Weiss,~A. \emph{Z. anorg. allg. Chem.} \textbf{1954}, \emph{276},
  95--112\relax
\mciteBstWouldAddEndPuncttrue
\mciteSetBstMidEndSepPunct{\mcitedefaultmidpunct}
{\mcitedefaultendpunct}{\mcitedefaultseppunct}\relax
\EndOfBibitem
\bibitem[Deer \latin{et~al.}(1992)Deer, Howie, and Zussman]{deer1992}
Deer,~W.~A.; Howie,~R.~A.; Zussman,~J. \emph{An Introduction to the
  Rock-forming Minerals, 2nd ed.}; Longman Scientific and Technical: Essex,
  1992; pp 1--696\relax
\mciteBstWouldAddEndPuncttrue
\mciteSetBstMidEndSepPunct{\mcitedefaultmidpunct}
{\mcitedefaultendpunct}{\mcitedefaultseppunct}\relax
\EndOfBibitem
\bibitem[Zhang \latin{et~al.}(2015)Zhang, Zhou, Wang, Chen, Kawazoe, and
  Jena]{ZhangPNAS}
Zhang,~S.; Zhou,~J.; Wang,~Q.; Chen,~X.; Kawazoe,~Y.; Jena,~P. \emph{Proc.
  Natl. Acad. Sci. U. S. A.} \textbf{2015}, \emph{112}, 2372--2377\relax
\mciteBstWouldAddEndPuncttrue
\mciteSetBstMidEndSepPunct{\mcitedefaultmidpunct}
{\mcitedefaultendpunct}{\mcitedefaultseppunct}\relax
\EndOfBibitem
\bibitem[Franklin(2015)]{Franklin}
Franklin,~A.~D. \emph{Science} \textbf{2015}, \emph{349}, aab2750\relax
\mciteBstWouldAddEndPuncttrue
\mciteSetBstMidEndSepPunct{\mcitedefaultmidpunct}
{\mcitedefaultendpunct}{\mcitedefaultseppunct}\relax
\EndOfBibitem
\bibitem[Bonaccorso \latin{et~al.}(2015)Bonaccorso, Colombo, Yu, Stoller,
  Tozzini, Ferrari, Ruoff, and Pellegrini]{bonaccorso}
Bonaccorso,~F.; Colombo,~L.; Yu,~G.; Stoller,~M.; Tozzini,~V.; Ferrari,~A.~C.;
  Ruoff,~R.~S.; Pellegrini,~V. \emph{Science} \textbf{2015}, \emph{347},
  1246501\relax
\mciteBstWouldAddEndPuncttrue
\mciteSetBstMidEndSepPunct{\mcitedefaultmidpunct}
{\mcitedefaultendpunct}{\mcitedefaultseppunct}\relax
\EndOfBibitem
\bibitem[Landau and Lifshitz()Landau, and Lifshitz]{Landau}
Landau,~L.~D.; Lifshitz,~E.~M. \emph{Theory of Elasticity;} Pergamon: Oxford,
  1995\relax
\mciteBstWouldAddEndPuncttrue
\mciteSetBstMidEndSepPunct{\mcitedefaultmidpunct}
{\mcitedefaultendpunct}{\mcitedefaultseppunct}\relax
\EndOfBibitem
\bibitem[Lakes(1987)]{Lakes}
Lakes,~R. \emph{Science} \textbf{1987}, \emph{235}, 1038--1040\relax
\mciteBstWouldAddEndPuncttrue
\mciteSetBstMidEndSepPunct{\mcitedefaultmidpunct}
{\mcitedefaultendpunct}{\mcitedefaultseppunct}\relax
\EndOfBibitem
\bibitem[Jiang \latin{et~al.}(2016)Jiang, Chang, Guo, and Park]{JiangG1}
Jiang,~J.-W.; Chang,~T.; Guo,~X.; Park,~H.~S. \emph{Nano Lett.} \textbf{2016},
  \emph{16}, 5286--5290\relax
\mciteBstWouldAddEndPuncttrue
\mciteSetBstMidEndSepPunct{\mcitedefaultmidpunct}
{\mcitedefaultendpunct}{\mcitedefaultseppunct}\relax
\EndOfBibitem
\bibitem[Jiang and Park(2016)Jiang, and Park]{JiangG2}
Jiang,~J.-W.; Park,~H.~S. \emph{Nano Lett.} \textbf{2016}, \emph{16},
  2657--2662\relax
\mciteBstWouldAddEndPuncttrue
\mciteSetBstMidEndSepPunct{\mcitedefaultmidpunct}
{\mcitedefaultendpunct}{\mcitedefaultseppunct}\relax
\EndOfBibitem
\bibitem[Jiang and Park(2014)Jiang, and Park]{Jiang}
Jiang,~J.-W.; Park,~H.~S. \emph{Nat. Commun.} \textbf{2014}, \emph{5},
  4727\relax
\mciteBstWouldAddEndPuncttrue
\mciteSetBstMidEndSepPunct{\mcitedefaultmidpunct}
{\mcitedefaultendpunct}{\mcitedefaultseppunct}\relax
\EndOfBibitem
\bibitem[Du \latin{et~al.}(2016)Du, Maassen, Wu, Luo, Xu, and Ye]{DuBP}
Du,~Y.; Maassen,~J.; Wu,~W.; Luo,~Z.; Xu,~X.; Ye,~P.~D. \emph{Nano Lett.}
  \textbf{2016}, \emph{16}, 6701--6708\relax
\mciteBstWouldAddEndPuncttrue
\mciteSetBstMidEndSepPunct{\mcitedefaultmidpunct}
{\mcitedefaultendpunct}{\mcitedefaultseppunct}\relax
\EndOfBibitem
\bibitem[Jiang \latin{et~al.}(2016)Jiang, Kim, and Park]{jiang2016auxetic}
Jiang,~J.-W.; Kim,~S.~Y.; Park,~H.~S. \emph{Appl. Phys. Rev.} \textbf{2016},
  \emph{3}, 041101\relax
\mciteBstWouldAddEndPuncttrue
\mciteSetBstMidEndSepPunct{\mcitedefaultmidpunct}
{\mcitedefaultendpunct}{\mcitedefaultseppunct}\relax
\EndOfBibitem
\bibitem[{\"O}z{\c{c}}elik \latin{et~al.}(2014){\"O}z{\c{c}}elik, Cahangirov,
  and Ciraci]{ozccelik2014stable}
{\"O}z{\c{c}}elik,~V.~O.; Cahangirov,~S.; Ciraci,~S. \emph{Phys. Rev. Lett.}
  \textbf{2014}, \emph{112}, 246803\relax
\mciteBstWouldAddEndPuncttrue
\mciteSetBstMidEndSepPunct{\mcitedefaultmidpunct}
{\mcitedefaultendpunct}{\mcitedefaultseppunct}\relax
\EndOfBibitem
\bibitem[Wang \latin{et~al.}(2016)Wang, Li, Li, and Chen]{wang2016semi}
Wang,~Y.; Li,~F.; Li,~Y.; Chen,~Z. \emph{Nat. Commun.} \textbf{2016},
  \emph{7}\relax
\mciteBstWouldAddEndPuncttrue
\mciteSetBstMidEndSepPunct{\mcitedefaultmidpunct}
{\mcitedefaultendpunct}{\mcitedefaultseppunct}\relax
\EndOfBibitem
\bibitem[Baughman \latin{et~al.}(1998)Baughman, Shacklette, Zakhidov, and
  Stafstr{\"o}m]{baughman1998}
Baughman,~R.~H.; Shacklette,~J.~M.; Zakhidov,~A.~A.; Stafstr{\"o}m,~S.
  \emph{Nature} \textbf{1998}, \emph{392}, 362--365\relax
\mciteBstWouldAddEndPuncttrue
\mciteSetBstMidEndSepPunct{\mcitedefaultmidpunct}
{\mcitedefaultendpunct}{\mcitedefaultseppunct}\relax
\EndOfBibitem
\bibitem[Grima and Evans(2006)Grima, and Evans]{grima2006}
Grima,~J.~N.; Evans,~K.~E. \emph{J Mater Sci} \textbf{2006}, \emph{41},
  3193--3196\relax
\mciteBstWouldAddEndPuncttrue
\mciteSetBstMidEndSepPunct{\mcitedefaultmidpunct}
{\mcitedefaultendpunct}{\mcitedefaultseppunct}\relax
\EndOfBibitem
\bibitem[Liu((DSTO, Defence Science and Technology Organisation,
  2006))]{liu2006auxetic}
Liu,~Q. \emph{Literature review: materials with negative Poisson's ratios and
  potential applications to aerospace and defence}; (DSTO, Defence Science and
  Technology Organisation, 2006)\relax
\mciteBstWouldAddEndPuncttrue
\mciteSetBstMidEndSepPunct{\mcitedefaultmidpunct}
{\mcitedefaultendpunct}{\mcitedefaultseppunct}\relax
\EndOfBibitem
\bibitem[Wei \latin{et~al.}(2009)Wei, Fragneaud, Marianetti, and Kysar]{WeiPRB}
Wei,~X.; Fragneaud,~B.; Marianetti,~C.~A.; Kysar,~J.~W. \emph{Phys. Rev. B}
  \textbf{2009}, \emph{80}, 205407\relax
\mciteBstWouldAddEndPuncttrue
\mciteSetBstMidEndSepPunct{\mcitedefaultmidpunct}
{\mcitedefaultendpunct}{\mcitedefaultseppunct}\relax
\EndOfBibitem
\bibitem[S\'anchez-Portal \latin{et~al.}(1999)S\'anchez-Portal, Artacho, Soler,
  Rubio, and Ordej\'on]{DanielPRB}
S\'anchez-Portal,~D.; Artacho,~E.; Soler,~J.~M.; Rubio,~A.; Ordej\'on,~P.
  \emph{Phys. Rev. B} \textbf{1999}, \emph{59}, 12678--12688\relax
\mciteBstWouldAddEndPuncttrue
\mciteSetBstMidEndSepPunct{\mcitedefaultmidpunct}
{\mcitedefaultendpunct}{\mcitedefaultseppunct}\relax
\EndOfBibitem
\bibitem[Wei and Peng(2014)Wei, and Peng]{Qunwei}
Wei,~Q.; Peng,~X. \emph{Appl. Phys. Lett.} \textbf{2014}, \emph{104},
  251915\relax
\mciteBstWouldAddEndPuncttrue
\mciteSetBstMidEndSepPunct{\mcitedefaultmidpunct}
{\mcitedefaultendpunct}{\mcitedefaultseppunct}\relax
\EndOfBibitem
\bibitem[Qiao \latin{et~al.}(2014)Qiao, Kong, Hu, Yang, and Ji]{JiweiNatcom}
Qiao,~J.; Kong,~X.; Hu,~Z.-X.; Yang,~F.; Ji,~W. \emph{Nat. Commun.}
  \textbf{2014}, \emph{5}, 4475\relax
\mciteBstWouldAddEndPuncttrue
\mciteSetBstMidEndSepPunct{\mcitedefaultmidpunct}
{\mcitedefaultendpunct}{\mcitedefaultseppunct}\relax
\EndOfBibitem
\bibitem[Keskar and Chelikowsky(1992)Keskar, and Chelikowsky]{Keskar}
Keskar,~N.~R.; Chelikowsky,~J.~R. \emph{Nature} \textbf{1992}, \emph{358}, 222
  -- 224\relax
\mciteBstWouldAddEndPuncttrue
\mciteSetBstMidEndSepPunct{\mcitedefaultmidpunct}
{\mcitedefaultendpunct}{\mcitedefaultseppunct}\relax
\EndOfBibitem
\bibitem[Wu \latin{et~al.}(2016)Wu, Varshney, Lee, Zhang, Wohlwend, Roy, and
  Luo]{XufeiNL}
Wu,~X.; Varshney,~V.; Lee,~J.; Zhang,~T.; Wohlwend,~J.~L.; Roy,~A.~K.; Luo,~T.
  \emph{Nano Lett.} \textbf{2016}, \emph{16}, 3925--3935\relax
\mciteBstWouldAddEndPuncttrue
\mciteSetBstMidEndSepPunct{\mcitedefaultmidpunct}
{\mcitedefaultendpunct}{\mcitedefaultseppunct}\relax
\EndOfBibitem
\bibitem[Wu \latin{et~al.}(2016)Wu, Varshney, Lee, Pang, Roy, and
  Luo]{XufeiarXiv}
Wu,~X.; Varshney,~V.; Lee,~J.; Pang,~Y.; Roy,~A.~K.; Luo,~T. \emph{arXiv
  preprint arXiv:1607.06542} \textbf{2016}, \relax
\mciteBstWouldAddEndPunctfalse
\mciteSetBstMidEndSepPunct{\mcitedefaultmidpunct}
{}{\mcitedefaultseppunct}\relax
\EndOfBibitem
\bibitem[Zubov and Firsova(1962)Zubov, and Firsova]{Zubov}
Zubov,~V.; Firsova,~M. \emph{Soviet. Phys. Cryst.} \textbf{1962}, \emph{7},
  374--376\relax
\mciteBstWouldAddEndPuncttrue
\mciteSetBstMidEndSepPunct{\mcitedefaultmidpunct}
{\mcitedefaultendpunct}{\mcitedefaultseppunct}\relax
\EndOfBibitem
\bibitem[Ohno(1995)]{Ohno}
Ohno,~I. \emph{J. Phys. Earth} \textbf{1995}, \emph{43}, 157--169\relax
\mciteBstWouldAddEndPuncttrue
\mciteSetBstMidEndSepPunct{\mcitedefaultmidpunct}
{\mcitedefaultendpunct}{\mcitedefaultseppunct}\relax
\EndOfBibitem
\bibitem[Kimizuka \latin{et~al.}(2007)Kimizuka, Ogata, Li, and Shibutani]{Kim}
Kimizuka,~H.; Ogata,~S.; Li,~J.; Shibutani,~Y. \emph{Phys. Rev. B}
  \textbf{2007}, \emph{75}, 054109\relax
\mciteBstWouldAddEndPuncttrue
\mciteSetBstMidEndSepPunct{\mcitedefaultmidpunct}
{\mcitedefaultendpunct}{\mcitedefaultseppunct}\relax
\EndOfBibitem
\bibitem[Andrew \latin{et~al.}(2012)Andrew, Mapasha, Ukpong, and
  Chetty]{Andrew}
Andrew,~R.~C.; Mapasha,~R.~E.; Ukpong,~A.~M.; Chetty,~N. \emph{Phys. Rev. B}
  \textbf{2012}, \emph{85}, 125428\relax
\mciteBstWouldAddEndPuncttrue
\mciteSetBstMidEndSepPunct{\mcitedefaultmidpunct}
{\mcitedefaultendpunct}{\mcitedefaultseppunct}\relax
\EndOfBibitem
\end{mcitethebibliography}

\providecommand{\latin}[1]{#1}
\makeatletter
\providecommand{\doi}
  {\begingroup\let\do\@makeother\dospecials
  \catcode`\{=1 \catcode`\}=2 \doi@aux}
\providecommand{\doi@aux}[1]{\endgroup\texttt{#1}}
\makeatother
\providecommand*\mcitethebibliography{\thebibliography}
\csname @ifundefined\endcsname{endmcitethebibliography}
  {\let\endmcitethebibliography\endthebibliography}{}

\end{document}